\documentclass[reprint, longbibliography]{revtex4-1}

\usepackage[utf8]{inputenc}

\usepackage[lining,semibold]{libertine} 
\usepackage{amsthm}
\usepackage[libertine, cmintegrals, bigdelims, vvarbb]{newtxmath}

\usepackage{amsmath}
\usepackage{amsfonts}
\usepackage{mathrsfs}
\usepackage{gensymb}
\usepackage{bbm}
\usepackage{dsfont}

\usepackage{kbordermatrix}
\renewcommand{\kbldelim}{(}
\renewcommand{\kbrdelim}{)}

\usepackage{chemformula}
\usepackage[caption=false]{subfig}

\usepackage{soul}
\usepackage{xcolor}

\theoremstyle{definition}

\usepackage{scalerel} 

\definecolor{webgreen}{rgb}{0,.5,0}
\definecolor{webbrown}{rgb}{.6,0,0}
\definecolor{grigio}{rgb}{.85,.85,.85} 
\definecolor{RoyalBlue}{rgb}{0.0, 0.14, 0.4}
\definecolor{skyblue1}{rgb}{0.45,0.62,0.81}
\definecolor{skyblue2}{rgb}{0.2,0.39,0.64}
\definecolor{skyblue3}{rgb}{0.13,0.29,0.53}
\definecolor{scarlet1}{rgb}{0.93,0.16,0.16}
\definecolor{scarlet2}{rgb}{0.8,0,0}
\definecolor{scarlet3}{rgb}{0.64,0,0}

\definecolor{g}{gray}{0.50}

\usepackage{hyperref}
\hypersetup{%
    colorlinks=true, linktocpage=true, pdfstartpage=1, pdfstartview=FitV,%
    breaklinks=true, pdfpagemode=UseNone, pageanchor=true, pdfpagemode=UseOutlines,%
    plainpages=false, bookmarksnumbered, bookmarksopen=true, bookmarksopenlevel=1,%
    hypertexnames=true, pdfhighlight=/O,
    urlcolor=webbrown, linkcolor=RoyalBlue, citecolor=webgreen, 
    pdftitle={},%
    pdfauthor={Francesco Avanzini},%
    pdfsubject={},%
    pdfkeywords={},%
    pdfcreator={pdfLaTeX},%
    pdfproducer={LaTeX REVTeX}%
}
\newcommand{\norm}[1]{\lVert#1\rVert}

\newcommand{\steady}[1]{\overline{#1}}

\newcommand{\dt}{\mathrm d_t}
\newcommand{\ddt}{\frac{\mathrm d}{\mathrm dt}}
\newcommand{\dddt}[1]{{\mathrm d#1}/{\mathrm dt}}

\newcommand{\eq}{\text{eq}}

\newcommand{\muteindex}{i}

\newcommand{\one}{\mathbb 1}

\newcommand{\chemspecies}{\alpha}
\newcommand{\chemostat}{{\alpha_c}}
\newcommand{\chemspeciesvec}{{\boldsymbol \chemspecies}}

\newcommand{\conc}{{\boldsymbol z}}
\newcommand{\setinternal}{X}
\newcommand{\setexchanged}{Y}
\newcommand{\setpotential}{{\setexchanged_p}}
\newcommand{\setforce}{{\setexchanged_f}}
\newcommand{\intconc}{\boldsymbol x}
\newcommand{\exconc}{\boldsymbol y}

\newcommand{\moieties}{\boldsymbol m}

\newcommand{\protocol}{{\boldsymbol \pi}}

\newcommand{\elrct}{\rho}

\newcommand{\curr}{{\boldsymbol j}}
\newcommand{\currel}{j}

\newcommand{\excurr}{{\boldsymbol I}}
\newcommand{\excurrY}{{\boldsymbol I}^{\setexchanged}}
\newcommand{\excurrel}{{ I}}

\newcommand{\stcoeff}[1]{\boldsymbol \nu_{#1\elrct}}

\newcommand{\matS}{{\mathbb S}}
\newcommand{\colS}{{\boldsymbol{ S}}_\elrct}

\newcommand{\matSX}{{\mathbb S}^{\setinternal}}
\newcommand{\matSY}{{\mathbb S}^{\setexchanged}}

\newcommand{\conslaw}{{\boldsymbol \ell}}
\newcommand{\conslawel}{ \ell}
\newcommand{\consquantity}{L}
\newcommand{\conslawindex}{\lambda}
\newcommand{\brokenindex}{{\lambda_b}}
\newcommand{\unbrokenindex}{{\lambda_u}}
\newcommand{\brokenconsquantity}{{\boldsymbol L}^{b}}

\newcommand{\matLb}{\mathbb L^{{b}}}

\newcommand{\matLbYp}{\mathbb L^{{b}}_{\setpotential}}
\newcommand{\matLbYf}{\mathbb L^{{b}}_{\setforce}}

\newcommand{\chempotential}{\mu}
\newcommand{\vecchempotential}{{\boldsymbol \mu}}
\newcommand{\stchempotential}{\mu^{\circ}}
\newcommand{\epr}{\dot{\Sigma}}

\newcommand{\semigrand}{\mathcal G}

\newcommand{\ncwr}{\dot{w}_{\text{nc}}}
\newcommand{\ssncwr}{\steady{\dot{w}}_{\text{nc}}}
\newcommand{\cdwr}{\dot{w}_{\text{driv}}}

\newcommand{\ncforce}{{\boldsymbol{\mathcal F}}{}_\setforce}

\newcommand{\concin}{\conc_0}
\newcommand{\intconcin}{\intconc{}_0}
\newcommand{\exconcin}{\exconc{}_0}

\newcommand{\excurrYa}{{\boldsymbol I}^{\setexchanged}_{ 1}}
\newcommand{\conca}{\conc_1}
\newcommand{\intconca}{\intconc{}_1}
\newcommand{\exconca}{\exconc{}_1}
\newcommand{\excurrYb}{{\boldsymbol I}^{\setexchanged}_{ 2}}

\newcommand{\intconcb}{\intconc{}_2}
\newcommand{\exconcb}{\exconc{}_2}
\newcommand{\protocolb}{\protocol_2}

\newcommand{\exfluxes}{fluxes}

\makeatletter
\def\maketag@@@#1{\hbox{\m@th\normalfont\normalsize#1}}
\makeatother

\DeclareMathAlphabet{\mathpzc}{OT1}{pzc}{m}{it}

\begin{document}
\title{Thermodynamics of Concentration vs Flux Control in Chemical Reaction Networks}

\newcommand\unilu{\affiliation{Complex Systems and Statistical Mechanics, Department of Physics and Materials Science, University of Luxembourg, L-1511 Luxembourg}}
\author{Francesco Avanzini}
\email{francesco.avanzini@uni.lu}
\unilu
\author{Massimiliano Esposito}
\email{massimiliano.esposito@uni.lu}
\unilu

\date{\today}

\begin{abstract}
We investigate the thermodynamic implications of two control mechanisms of open chemical reaction networks.
The first controls the concentrations of the species that are exchanged with the surroundings, while the other controls the exchange fluxes. 
We show that the two mechanisms can be mapped one into the other and that the thermodynamic theories usually developed in the framework of concentration control can be applied to flux control as well.
This implies that the thermodynamic potential and the fundamental forces driving chemical reaction networks out of equilibrium can be identified in the same way for both mechanisms.
By analyzing the dynamics and thermodynamics of a simple enzymatic model we also show that, while the two mechanisms are equivalent at steady state, the flux control may lead to fundamentally different regimes where systems achieve stationary growth.
\end{abstract}

\maketitle



\section{Introduction}
Stochastic thermodynamics~\cite{Sekimoto2010, Jarzynski2011, Seifert2012, VanDenBroeck2015} provides a rigorous ground to characterize the energetics of systems driven arbitrarily far from equilibrium.
It has been formulated for open chemical reaction networks (CRNs) undergoing a stochastic~\cite{Gaspard2004, Schmiedl2007, Rao2018b} or a deterministic dynamics~\cite{Qian2005, Polettini2014, Rao2016}.
Crucially, the latter formulation is equivalent
to the former in macroscopic limit~\cite{Ge2016a, Ge2016b, Rao2016, Rao2018b}, where fluctuations become negligible and the stochastic dynamics converges to the deterministic one~\cite{Kurtz1970,Kurtz1971,Kurtz1972},
and to Gibbs' chemical thermodynamics at equilibrium.
In these frameworks, CRNs are driven out of equilibrium by the exchanges of some chemical species with the surroundings which are in general represented in terms of infinitely large (but also finite~\cite{Fritz2020}) reservoirs, called chemostats, controlling the concentrations of the exchanged species.
This theory has been used to quantify the energetic cost for creating patterns~\cite{Falasco2018a} and waves~\cite{ Avanzini2019a}, and sustaining the growth process of macromolecules, like copolymers~\cite{Andrieux2008, Blokhuis2017, Gaspard2020a} and biomolecules~\cite{Rao2015}.
Its connections to information geometry~\cite{Yoshimura2021a}, as well as to thermodynamic uncertainty relations and speed limits~\cite{Yoshimura2021b}, have been recently investigated.
Furthermore, the topological properties of CRNs, and in particular the conservation laws, have been exploited to decompose the entropy production rate in such a way as to identify the proper thermodynamic potential of open CRNs as well as the fundamental forces and chemostats breaking the detailed balance condition and driving CRNs out of equilibrium~\cite{Rao2018b, Avanzini2021}.

Controlling the concentrations of some species via chemostats is not the only way the surroundings can affect CRNs.
They can also control the exchange fluxes.
For instance, if the mitochondrial metabolism is analyzed as an open CRN, its surroundings correspond to the cytosolic processes
which do not control the concentrations of some species inside the mitochondrion, but instead they determine its exchange fluxes~\cite{Magnus1997,Magnus1998,Fall2001}.
Furthermore, in industrial continuous-flow stirred tank reactors~\cite{Aris1989} as well as in biotechnological continuous cultures~\cite{Stockar2013},  systems are fed by reactants at a constant rate while some other species are continuously extracted with a controlled outflow.
Hence, recent studies started considering flux control mechanisms to drive CRNs out of equilibrium rather than concentration control via chemostats.
Some of these studies focused on continuous-flow stirred tank reactors which have been analyzed from a kinetic~\cite{Craciun2005} as well as a thermodynamic~\cite{Blokhuis2018} standpoint. 
This formulation has for example been used to study the emergence of homochirality~\cite{Laurente2021a, Laurente2021b} and reformulate the thermodynamic evolution theorem in modern terms~\cite{Hochberg2020}.
Other studies considered CRNs under both concentration and flux control, and investigated how the sensitivity of CRNs to changes in the extraction fluxes can be adjusted by nonequilibrium futile cycles~\cite{Qian2006}.

In this work, we provide a systematic comparison between open CRNs under concentration control and flux control from a thermodynamic standpoint. 
We focus on deterministic CRNs whose dynamical description is summarized in Sec.~\ref{sec:dyn}.
In Sec.~\ref{sec:conc}, we discuss concentration control and its thermodynamic characterization, in particular the decomposition of the entropy production rate in terms of thermodynamic potential and fundamental forces.
We turn to flux control in Sec.~\ref{sec:flux} where we show that it can be mapped into a specific protocol of concentration control. 
This proves that the same thermodynamic theories developed for CRNs under concentration control can be used for CRNs under flux control, including the decomposition of the entropy production rate.
In Sec.~\ref{sec:example}, we investigate the different thermodynamic properties of the two control mechanisms by considering a model system of an enzymatic reaction.
We focus on how the entropy production rate, the thermodynamic potential and the work done by the fundamental forces evolve depending on the control mechanism.
General conclusions are drawn in Sec.~\ref{sec:discussion} where we also discuss the implications of our results which should be particularly relevant to characterize the energetics of different growth regimes in autocatalytic systems.


\section{Dynamics of Open CRNs\label{sec:dyn}}

CRNs are considered here as ideal dilute solutions composed of chemical species $\chemspeciesvec =(\dots, \chemspecies, \dots)^\intercal$ which interconvert via chemical reactions.
In the macroscopic limit, the evolution of the abundances of the chemical species, expressed in terms of concentrations $\conc =(\dots,[\chemspecies] ,\dots)^\intercal$, follows a deterministic rate equation:
\begin{equation}
\ddt\conc=\matS\curr+\excurr\,.
\label{eq:rate}
\end{equation}

The first term on the r.h.s. of Eq.~\eqref{eq:rate}, i.e., $\matS\curr$, accounts for the variation of the concentrations due to the chemical reactions.
Each column $\colS$ of the stoichiometric matrix $\matS$ encodes the net variations in terms of number of molecules of each species undergoing  (the elementary~\cite{Svehla1993}) reaction $\elrct$,
\begin{equation}
\chemspeciesvec\cdot \stcoeff{+}  \ch{<=>[ $+\elrct$ ][ $-\elrct$ ]} \chemspeciesvec\cdot \stcoeff{-}\,,
\label{eq:chem_rct}
\end{equation}
where $\stcoeff{+}$ (resp. $\stcoeff{-}$) is the vector of the stoichiometric coefficients of the forward (resp. backward) reaction.
Thus, $\colS= \stcoeff{-} - \stcoeff{+}$.
Each entry $\currel_\elrct$ of the vector $\curr = (\dots,\currel_\elrct,\dots)^\intercal$ specifies the net current of every reaction $\elrct$ as the difference between the forward and backward one, i.e., $\currel_\elrct = \currel_{+\elrct} - \currel_{-\elrct}$, which satisfy mass-action kinetics: 
\begin{equation}
\currel_{\pm\elrct} = k_{\pm\elrct}\conc^{\stcoeff{\pm}}\,,
\label{eq:massaction}
\end{equation}
with $k_{\pm\elrct}$ the kinetic constants of the forward/backward reaction and ${\boldsymbol a}^{\boldsymbol b} = \prod_\muteindex a_\muteindex^{b_\muteindex}$.

The second term on the r.h.s. of Eq.~\eqref{eq:rate} is the exchange vector $\excurr = (\dots, \excurrel_\chemspecies,\dots)^\intercal$ accounting for net matter flux of each species between the CRN and the surroundings independently of the control mechanism.
It vanishes if CRNs are closed.
In open CRNs, we split the chemical species into 
the internal species $\chemspecies \in \setinternal$, namely those that are not exchanged (i.e., $\excurrel_\chemspecies = 0$), 
and the exchanged species $\chemspecies \in \setexchanged$.
By applying the same splitting to the concentration vector $\conc = (\intconc, \exconc)$ and the stoichiometric matrix,
\begin{equation}
\matS=\begin{pmatrix}
\matSX \\ 
\matSY\\
\end{pmatrix}\,,
\end{equation}
the rate equation~\eqref{eq:rate} can be specialized as
\begin{align}
&\ddt \intconc=\matSX \curr \label{eq:rateX}\,,\\
&\ddt \exconc=\matSY \curr+ \excurrY\label{eq:rateY}\,,
\end{align}
with $\excurrY = (\dots, \excurrel_\chemspecies,\dots)^\intercal_{\chemspecies\in\setexchanged}$ collecting all the nonw null entries of $\excurr$.

Closed CRNs must be \textit{detailed balanced}, namely an equilibrium steady state $\conc_\eq$ such that $\curr(\conc_\eq)=0$ always exists.
In open CRNs, the detailed balanced condition can be broken because of the exchanges of the chemical species.


\section{Concentration Control\label{sec:conc}}
A first way in which the matter exchanges with the surroundings can affect the dynamics of CRNs is by controlling the concentrations of the $\setexchanged$ species. 
Thus, $\exconc$ cease to be dynamical variables, unlike the concentrations of the internal species $\intconc$, and become externally controlled parameters, while Eq.~\eqref{eq:rateY} becomes  merely a definition of the exchange \exfluxes\text{ }$\excurrY$.

We specialize this concentration control in two classes.
The first one maintains the concentrations $\exconc$ constant, i.e., $\dddt{\exconc}=0$,
and the corresponding exchange \exfluxes\text{ }are given by
\begin{equation}
\excurrY = - \matSY \curr\,.
\label{eq:excurr_constconccontrol}
\end{equation}
In this case, CRNs are said to be \textit{autonomous} since the control parameters $\exconc$ have fixed values. 
The second one forces the concentrations $\exconc$ to follow a specific protocol, i.e., ${\exconc}=\protocol$,
and the corresponding exchange \exfluxes\text{ }must satisfy
\begin{equation}
\excurrY = \ddt\protocol - \matSY \curr\,.
\end{equation}
In this case, CRNs are said to be \textit{nonautonomous}.

\begin{figure}[t]
  \centering
  \includegraphics[width=0.99\columnwidth]{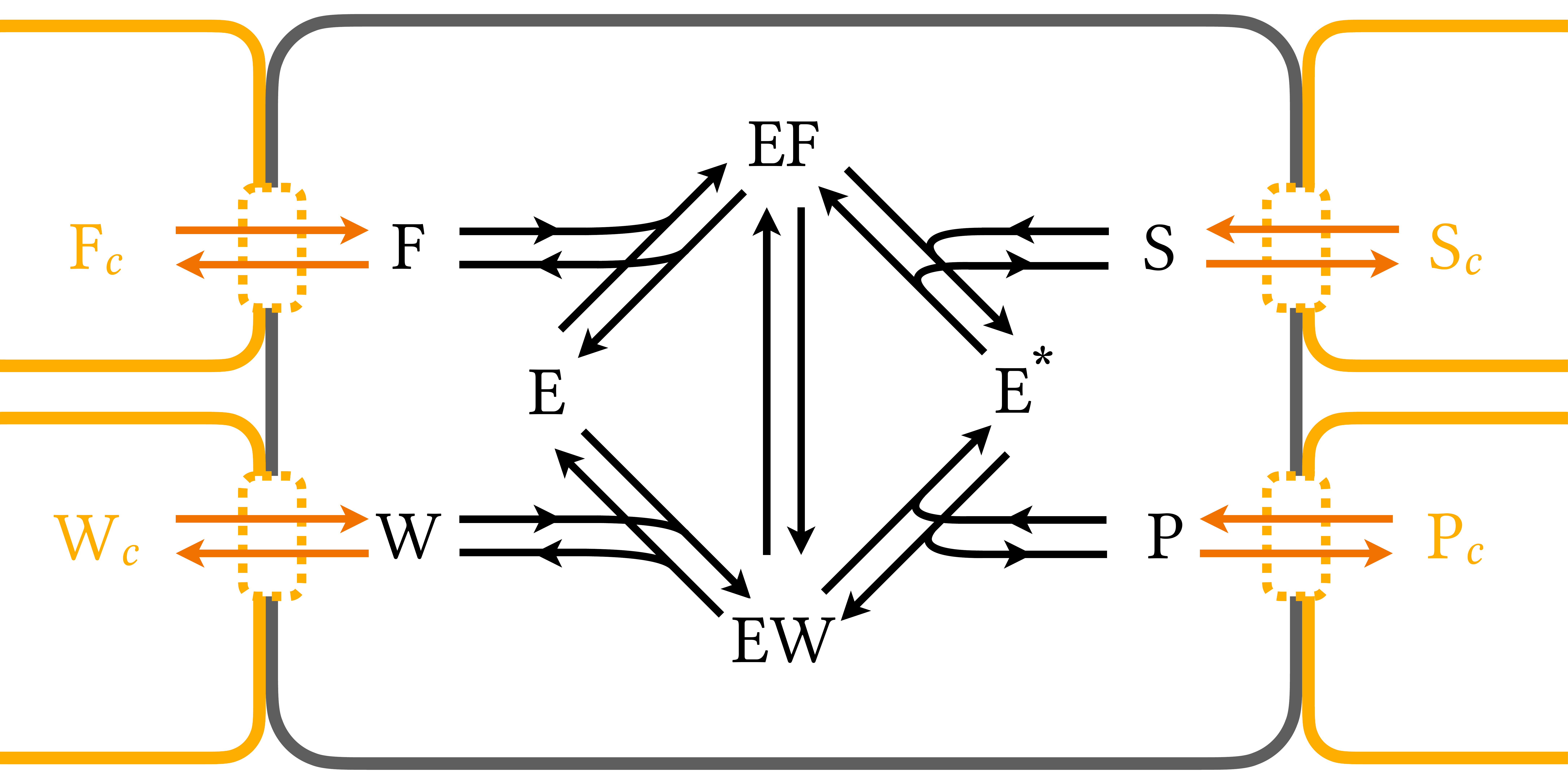}
\caption{Pictorial illustration of the CRN~\eqref{eq:example} when subjected to (autonomous) concentration control via chemostats fixing the concentrations of \ch{F}, \ch{S}, \ch{W}, \ch{P}.}
\label{fig:chemostats}
\end{figure}

The concentration control is the most common one in the framework of nonequilibrium thermodynamics.
Physically, it corresponds to coupling a CRN with external reservoirs called \textit{chemostats} as in Fig.~\ref{fig:chemostats}.
Each chemostat exchanges a specific $\setexchanged$ species according to the process
\begin{equation}
\chemspecies \ch{<=>[ $$ ][ $$ ]} \chemostat
\label{eq:ex_process_chemostat}
\end{equation}
that is assumed to always be at equilibrium.
From a thermodynamic standpoint, this implies that the chemical potential $\chempotential_\chemspecies$ of each $\chemspecies\in\setexchanged$ species, namely its contribution to the free energy, is controlled by the chemostat, i.e, $\chempotential_\chemspecies = \chempotential_\chemostat$.
The chemical potentials are given by 
\begin{equation}
\chempotential_\chemspecies = \stchempotential_\chemspecies + RT \ln[\chemspecies]\,,
\label{eq:chempot}
\end{equation}
where $\stchempotential_\chemspecies $ is the standard chemical potential of the species $\chemspecies$, $R$ is the gas constant, and $T$ is the temperature fixed by the solvent.
Consequently, the chemostats control the concentrations of the $\setexchanged$ species,
\begin{equation}
[\chemspecies] = \exp\Big(\frac{\chempotential_\chemostat - \stchempotential_\chemspecies}{RT}\Big)
\text{ }\text{ }\text{ }\forall\chemspecies\in\setexchanged\,,
\end{equation}
that are said to be \textit{chemostatted}. 
Depending on whether the chemical potentials $\chempotential_\chemostat$ are constant in time or not, the concentrations $\exconc$ have fixed values or not.
Note that in previous studies~\cite{Schmiedl2007,Polettini2014,Falasco2019a} the $\setexchanged$ species are directly regarded as chemostats \textit{within} the CRN.
These two interpretations are formally equivalent, but the former has the advantage of being easier to compare to the flux control as we will see.

Within this framework, the entropy production rate of elementary reactions,
\begin{equation}
\epr = R\sum_{\elrct} \big(\currel_{+\elrct} - \currel_{-\elrct} ) \ln\frac{\currel_{+\elrct} }{\currel_{-\elrct} },
\end{equation}
has been decomposed~\cite{Rao2018b, Avanzini2021} as
\begin{equation}
T\epr = -\ddt\semigrand +\ncwr + \cdwr\,,
\label{eq:epr_decomposition}
\end{equation}
where the \textit{semi-grand free energy} $\semigrand$ is the proper thermodynamic potential of open CRNs;
$\ncwr$ is the \textit{nonconservative work rate} accounting for the energetic cost of breaking detailed balance; 
$\cdwr$ is the \textit{driving work rate} quantifying the energetic cost of changing the equilibrium state to which CRNs would relax if they were detailed balance.

The derivation of Eq.~\eqref{eq:epr_decomposition} is discussed in detail in Ref.~\cite{Rao2018b} for stochastic and in Ref.~\cite{Avanzini2021} for deterministic CRNs. 
Here, we recap the general strategy and we provide the explicit expressions of $\semigrand$, $\ncwr$, and  $\cdwr$.
To do so, we have to introduce the \textit{conservation laws} and the splitting of the chemostatted species $\setexchanged$ into potential $\setpotential$ and force species $\setforce$.
The conservation laws~\cite{Polettini2014,Rao2016} are linearly independent vectors $\{\conslaw^{\conslawindex}\}$ in the cokernel of the stoichiometric matrix, 
\begin{equation}
\conslaw^{\conslawindex}\cdot\matS =0\,.
\label{eq:conslaw}
\end{equation}
Indeed, all the scalars $\consquantity^\conslawindex \equiv \conslaw^{\conslawindex}\cdot \conc$ would be conserved quantities if CRNs were closed: $\dddt{\consquantity} = \conslaw^{\conslawindex}\cdot \matS \curr = 0$.
These scalars (or their linear combinations) quantify the concentrations of the so-called \textit{moieties},  i.e., parts of (or entire) molecules that remain intact in all the reactions.
Note that the total mass must be conserved  in closed CRNs and thus the set $\{\conslaw^{\conslawindex}\}$ is never empty.
In open CRNs, the set of conservation laws can be split into the \textit{unbroken} conservation laws $\{\conslaw^\unbrokenindex\}$ and the \textit{broken} conservation laws $\{\conslaw^\brokenindex\}$.
The former are the largest subset of conservation laws that can be written with null entries for the $\setexchanged$ species, i.e, $\conslawel^\unbrokenindex_\chemspecies =0$ $\forall \chemspecies \in \setexchanged$.
The scalar quantities $\consquantity^\unbrokenindex = \conslaw^\unbrokenindex\cdot\conc$ are thus conserved even if CRNs are opened: 
\begin{equation}
\ddt\consquantity^\unbrokenindex =
\underbrace{\conslaw^{\unbrokenindex}\cdot \matS \curr}_{ = 0} + \sum_{\chemspecies \in \setexchanged}\underbrace{\conslawel^\unbrokenindex_\chemspecies}_{=0} \excurrel_\chemspecies=0\,.
\end{equation}
The later are the other conservation laws and the scalar quantities $\consquantity^\brokenindex = \conslaw^\brokenindex\cdot\conc$ are in general not conserved when CRNs are opened, i.e.,
\begin{equation}
\ddt\consquantity^\brokenindex = 
\underbrace{\conslaw^{\brokenindex}\cdot \matS \curr}_{ = 0} + \sum_{\chemspecies \in \setexchanged}\underbrace{\conslawel^\brokenindex_\chemspecies}_{\neq0} \excurrel_\chemspecies\neq0\,,
\label{eq:broken_cq}
\end{equation}
because the corresponding moieties are exchanged with the chemostats.
Note that the total mass is not conserved  in open CRNs and thus the set $\{\conslaw^{\brokenindex}\}$ is never empty.

Chemostatting a species does not always break a conservation law~\cite{Rao2016,Rao2018b, Rao2018a, Falasco2018a}.
The potential species $\setpotential$ are those that break the conservation laws when chemostatted, while the force species are the other species $\setforce = \setexchanged\setminus \setpotential$.
On the one hand, every time a $\setpotential$ species is chemostatted a new moiety is exchanged between the CRN and the chemostats. 
The concentrations of the exchanged moieties are expressed in terms of linear combinations~\footnote{
The choice of the conservation laws~\eqref{eq:conslaw} is not unique, and different choices identify different moieties. 
The linear combination in Eq.~\eqref{eq:conc_moiety} selects a particular set of conservation laws and, correspondingly, of moieties such that each moiety is exchanged with only one chemostat as long as no $\setforce$ species are chemostatted:
$\dt\moieties = \excurr_\setpotential$}
of the broken conserved quantities~\eqref{eq:broken_cq} according to
\begin{equation}
\moieties = (\matLbYp)^{-1}\brokenconsquantity
\label{eq:conc_moiety}
\end{equation}
where $\brokenconsquantity = (\dots, \consquantity^\brokenindex, \dots)^\intercal$ and $\matLbYp$ is the (invertible) submatrix for the $\setpotential$ species of the matrix $\matLb$ whose rows are the broken conservation laws (i.e., $\matLb$ and $\matLbYp$ have $\{\conslawel^\brokenindex_\chemspecies\}$ and $\{\conslawel^\brokenindex_\chemspecies\}_{\chemspecies\in\setpotential}$ as entries, respectively).
On the other hand, when a $\setforce$ species is chemostatted, the corresponding chemostat exchanges a moiety with the CRN that is already exchanged with another one thus establishing a flux of a moiety between two (or more) chemostats~\footnote{
When also the $\setforce$ species are chemostatted, the concentration of each moiety in Eq.~\eqref{eq:conc_moiety}
depends on the flux with the corresponding $\setpotential$ chemostat and on the fluxes with the $\setforce$ chemostats that exchange the same moiety 
according to $\dt\moieties = \excurr_\setpotential + \big(\matLbYp\big)^{-1}\matLbYf \excurr_\setforce$ (with $\matLbYf$ the submatrix for the $\setforce$ species of $\matLb$)}.

The different role played by the $\setpotential$ and $\setforce$ specie provides a rigorous ground to identify the thermodynamic potential of open CRNs and to decompose the entropy production rate~\eqref{eq:epr_decomposition}.
As in equilibrium thermodynamics when passing from the canonical to the grand canonical ensemble, the thermodynamic potential $\semigrand$ is obtained from the Gibbs free energy $G$ by eliminating the energetic contributions of the matter exchanged with the particle  reservoir that in this context corresponds to the chemostats.
The latter accounts for the concentrations of the moieties~\eqref{eq:conc_moiety} times the chemical potential of the $\setpotential$ species.
Thus,
\begin{equation}
\semigrand = \underbrace{\vecchempotential \cdot \conc - RT \norm{\conc}}_{= \text{ }G} - \vecchempotential_\setpotential \cdot \moieties\,,
\label{eq:semigrand}
\end{equation}
with $\vecchempotential = (\dots,\chempotential_\chemspecies,\dots)^\intercal$, 
$\vecchempotential_\setpotential = (\dots,\chempotential_\chemspecies,\dots)^\intercal_{\chemspecies\in\setpotential}$, 
and $\norm{\conc}=\sum_\chemspecies [\chemspecies]$. 
Exploiting the conservation law, one can further verify that $\semigrand$ is lower bounded by its equilibrium value, i.e. $\semigrand\geq\semigrand_\eq$.

The nonconservative work rate,
\begin{equation}
\ncwr = \ncforce\cdot\excurr_\setforce\,,
\label{eq:ncwr}
\end{equation}
quantifies the energetic cost of sustaining fluxes of moieties between chemostats by means of the nonconservative force 
$\ncforce = \big( \vecchempotential_\setforce \cdot \one - \vecchempotential_\setpotential \cdot \big(\matLbYp\big)^{-1}\matLbYf \big)^\intercal$ (with $\matLbYf$ the submatrix for the $\setforce$ species of $\matLb$, i.e., the matrix with entries $\{\conslawel^\brokenindex_\chemspecies\}_{\chemspecies\in\setforce}$, and $\one$ the identity matrix).
When there are no $\setforce$ species, each moiety is exchanged with a single $\setpotential$ chemostat and there are no fluxes between chemostats.
Thus, $\ncwr$ vanishes. 
If there are $\setforce$ species, $\ncwr$ can still vanish if
\begin{equation}
\vecchempotential_\setforce = \big( \vecchempotential_\setpotential \cdot \big(\matLbYp\big)^{-1}\matLbYf \big)^\intercal\,.
\end{equation}  

The driving work rate,
\begin{equation}
\cdwr = -\Big(\ddt \vecchempotential_\setpotential \Big)\cdot \moieties \,,
\label{eq:cdwr}
\end{equation}
accounts for the time dependent manipulation of the chemical potentials of the moieties, and so of their concentrations, via the $\setpotential$ chemostats. 
Since the chemical potential of a species~\eqref{eq:chempot} is univocally determined by its concentration, the driving work rate obviously vanishes if $\dddt{\exconc} = 0$, namely under an autonomous concentration control.

We consider now three simple cases to illustrate the physical meaning of the thermodynamic quantities given in Eq.~\eqref{eq:semigrand},~\eqref{eq:ncwr} and~\eqref{eq:cdwr}.
First, when either no species are exchanged or only the concentrations of the $\setpotential$ species are maintained constant by the chemostats, Eq.~\eqref{eq:epr_decomposition} simplifies to $\dddt{\semigrand} = -T\epr \leq 0$.
This, together with the lower bound  $\semigrand\geq\semigrand_\eq$, implies that the CRN eventually reaches an equilibrium state after a nonequilibrium transient dynamics in which the semigrand free energy $\semigrand$ is dissipated, i.e., it is detailed balanced.
The particular equilibrium state to which the CRN relaxes depends on the concentrations of the $\setpotential$ species.
Second, when the concentrations of the $\setpotential$ are driven, the CRN is still detailed balanced, but the equilibrium state to which it would relax changes in time.
If the driving is fast enough, the CRN never reaches equilibrium and is thus maintained in a nonequilibrium transient dynamics.
In this case, Eq.~\eqref{eq:epr_decomposition} becomes $\dddt{\semigrand} = -T\epr + \cdwr$ and $\cdwr$ accounts for the energetic cost of driving the equilibrium state.
For this reason $\cdwr$ depends only on the $\setpotential$ species.
Third, when the concentrations of the $\setpotential$ species are maintained constant while also the $\setforce$ species are exchanged, detailed balance is broken:
the CRN is maintained out of equilibrium both in the transient dynamics and at steady state.
Indeed, Eq.~\eqref{eq:epr_decomposition} becomes $\dddt{\semigrand} = -T\epr + \ncwr$ which implies that the dissipation cannot vanish at steady state since $T\steady{\epr} = {\ssncwr}$.
Thus, $\ncwr$ quantifies the energetic cost of breaking detailed balance and $\ncforce$ is the nonconservative force preventing the CRN from reaching an equilibrium state.
%


\section{Flux Control\label{sec:flux}}
A second way in which the matter exchanges with the surroundings can affect the dynamics of CRNs is by controlling the exchange \exfluxes~$\excurrY$. 
The concentrations $\exconc$ are not controlled parameters, but they are still dynamical variables whose evolution follows Eq.~\eqref{eq:rateY}.

We can specialize this flux control in two classes.
In the first one, the exchange \exfluxes~$\excurrY$ are either constant or depend on the concentrations $\conc$. 
We refer to this regime as \textit{autonomous} since the control parameters~$\excurrY$ are not externally manipulated.
In the second one, the exchange \exfluxes~$\excurrY$ can be forced to follow an externally imposed protocol.
We refer to this regime as \textit{nonautonomous}.

\begin{figure}[t]
  \centering
  \includegraphics[width=0.99\columnwidth]{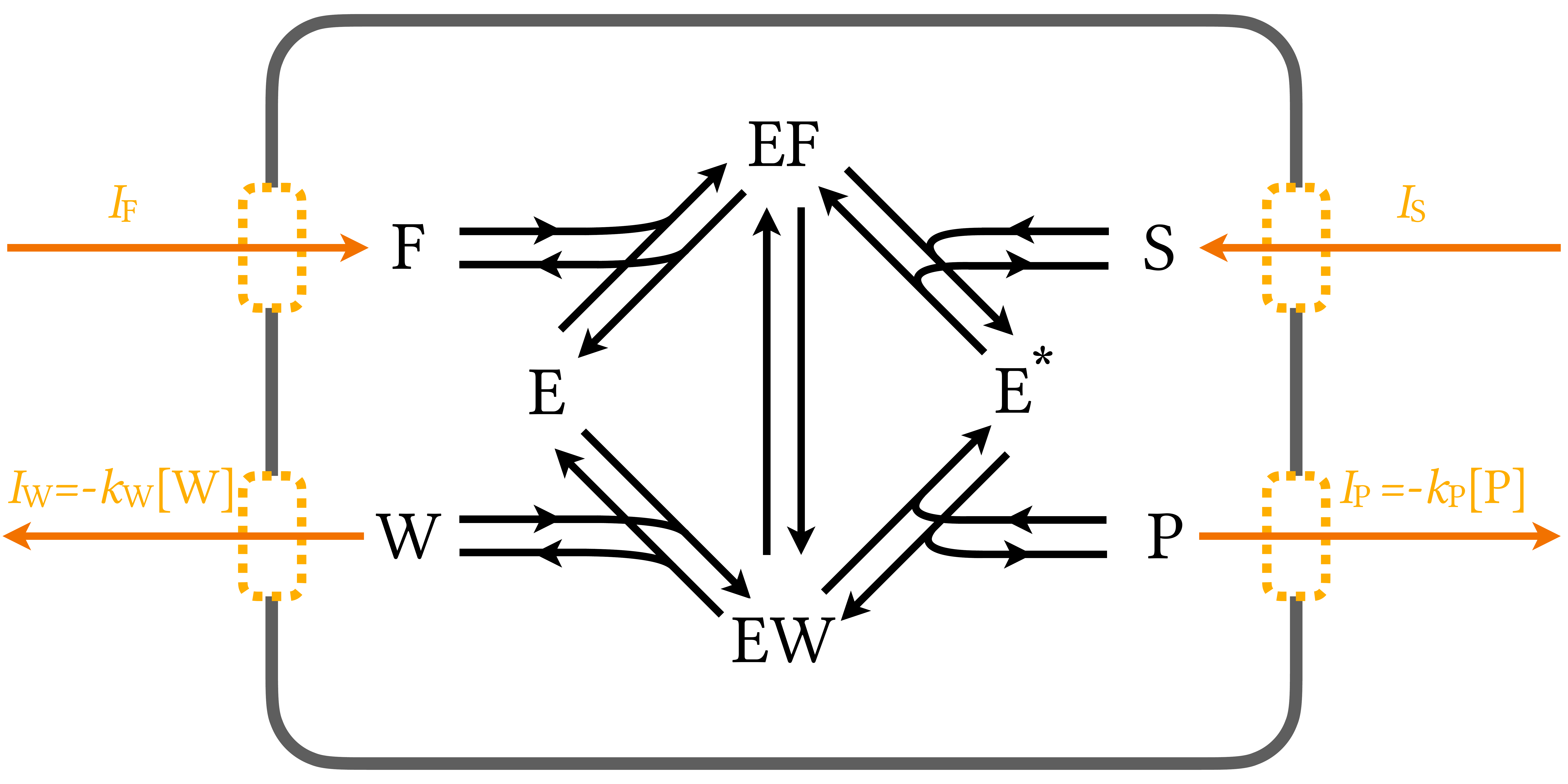}
\caption{Pictorial illustration of the CRN~\eqref{eq:example} when subjected to (autonomous) flux control. The species \ch{S} and \ch{F} are continuously injected at the constant rates $\excurrel_{\ch{S}}$ and $\excurrel_{\ch{F}}$, respectively. The species \ch{P} and \ch{W} are extracted at concentration-dependent rates $\excurrel_{\ch{P}} = -k_{\ch{P}}[\ch{P}]$ and $\excurrel_{\ch{W}} = -k_{\ch{W}} [\ch{W}]$, respectively.}
\label{fig:processes}
\end{figure}

The flux control physically corresponds to coupling a CRN with other external processes whose net effects on the concentrations of the $\setexchanged$ species are quantified by the exchange \exfluxes\text{ }$\excurrY$:
\begin{equation}
\chemspecies \ch{<->[ $\excurrel_\chemspecies$ ][ $$ ]} \dots\,,
\label{eq:ex_process_process}
\end{equation}
where we used the symbol $\ch{<->}$ to represent the fact that external processes can either inject or extract the species $\chemspecies\in\setexchanged$.
For example, some species can be injected in the CRN at a certain constant rate and extracted with an outflow that is proportional to their concentrations as in Fig.~\ref{fig:processes}.
This particular strategy is equivalent to a continuous-flow stirred tank reactor when applied to all the species~\cite{Blokhuis2018}.
Furthermore, chemostating can be considered as particular external processes~\eqref{eq:ex_process_chemostat} that fix the concentrations $\exconc$.

It is natural to wonder whether the decomposition of the entropy production rate~\eqref{eq:epr_decomposition} derived for CRNs coupled to chemostats can still be applied to this different framework or not.
We argue here that it can be by mapping the flux control into a nonautonomous concentration control.

Consider as first case a generic CRN where the exchange \exfluxes\text{ }$\excurrYa$ are controlled (here we use the subscript 1 to stress that these are the exchange \exfluxes\text{ }of the first case).
We do not make any assumption on the expressions of the entries of $\excurrYa$: they might be constant, or depend on the concentrations $\conc$ or follow an externally imposed protocol.
What matters is that the concentrations of the chemical species at a generic time~$t$,
\begin{equation}
\conca = (\intconca, \exconca)\,,
\end{equation}
are univocally determined by solving Eq.~\eqref{eq:rate} (or equivalently Eq.~\eqref{eq:rateX} and~\eqref{eq:rateY}) for given expressions of $\excurrYa$ and a given initial condition $\concin = (\intconcin, \exconcin)$.

Consider as second case the same CRN with the same initial condition, but where the concentrations of the $\setexchanged$ species are controlled.
If the concentrations $\exconcb$ follow an external protocol $\protocolb$ which is equivalent to their evolution in the first case, that is
\begin{equation}
\protocolb = \exconca
\text{ }\text{ }\text{ } \forall t\,,
\end{equation}
then also the evolution of the exchange \exfluxes\text{ }$\excurrYb$  and the internal concentrations will necessarily be the same for the whole dynamics:
\begin{equation}
\excurrYb = \excurrYa\,, \text{ }\text{ }\text{ }
\intconcb=\intconca\,,\text{ }\text{ }\text{ }
\forall t\,.
\end{equation}
This implies that the CRN cannot know whether it is subjected to a concentration control or a flux control:
it only experiences the net effects of the coupling with the surroundings, i.e., the exchange \exfluxes\text{ }$\excurrY$.
Therefore, the free energy balance between the CRN and the surroundings must be exactly the same in the two cases 
and, consequently, the decomposition of the entropy production in Eq.~\eqref{eq:epr_decomposition} still holds.
The difference between the two ways of controlling CRNs regards only their surroundings.
In particular, we emphasize that while chemostats are external reservoirs at equilibrium, the external processes can be out of equilibrium.

Furthermore, the derivation~\cite{Rao2018b,Avanzini2021} of the semigrand free energy~$\semigrand$, the nonconservative work rate~$\ncwr$ and the driving work rate~$\cdwr$ (summarized in Sec.~\ref{sec:conc}) is based only on which moieties (broken conserved quantities) are exchanged with the surroundings. 
It does not require any assumption on the kind of processes breaking the conservation laws.
Thus, $\semigrand$, $\ncwr$ and $\cdwr$ are given by Eq.~\eqref{eq:semigrand}, \eqref{eq:ncwr}, and \eqref{eq:cdwr}, respectively, whether CNRs are subjected to concentration control or flux control, and their physical meaning is exactly the same in both cases.


\section{Model System\label{sec:example}}
We now consider the following CRN,
\begin{equation}
\begin{split}
\ch{F + E &<=>[ ${+1}$ ][ ${-1}$ ] EF }\\
\ch{EF  &<=>[ ${+2}$ ][ ${-2}$ ] EW }\\
\ch{EW  &<=>[ ${+3}$ ][ ${-3}$ ] E + W }\\
\ch{S + EF &<=>[ ${+4}$ ][ ${-4}$ ] E^{*}}\\
\ch{ E^{*} &<=>[ ${+5}$ ][ ${-5}$ ] EW + P }
\end{split}
\label{eq:example}
\end{equation}
representing the active interconversion of the substrate \ch{S} into the product \ch{P} catalyzed by the enzyme \ch{E} and powered by the interconversion of the fuel \ch{F} into the waste \ch{W}. 
The species \ch{EF} (resp. \ch{EW}) is the complex enzyme-fuel (resp. enzyme-waste) while \ch{E^{*}} is the complex binding both the fuel and the substrate.
Given the stoichiometric matrix of the CRN~\eqref{eq:example},
\begin{equation}
{\matS}=
 \kbordermatrix{
    & \color{g}1 &\color{g}2&\color{g}3&\color{g}4&\color{g}5\cr
    \color{g}\ch{E} 	  &-1 &0 &1 &0&0\cr
    \color{g}\ch{EF}  	  &1 &-1&0&-1&0\cr
    \color{g}\ch{EW}  	  &0 &1&-1&0&1\cr
    \color{g}\ch{E^{*}}   &0 &0&0&1&-1\cr
    \color{g}\ch{P}        &0 &0&0&0&1\cr
    \color{g}\ch{W}       &0 &0&1&0&0\cr
    \color{g}\ch{S}	 &0 &0&0&-1&0\cr
    \color{g}\ch{F}   	&-1 &0&0&0&0\cr
  }\,,
  \label{eq:example_matS}
\end{equation}
we identify the conservation laws
\begin{equation}
 \conslaw^{\ch{E}}=
 \kbordermatrix{
     & \cr
    \color{g}\ch{E}    &1\cr
    \color{g}\ch{EF}   &1\cr
    \color{g}\ch{EW}   &1\cr
    \color{g}\ch{E^{*}}   &1\cr
    \color{g}\ch{P} &0\cr
    \color{g}\ch{W}   &0\cr
    \color{g}\ch{S} &0\cr
    \color{g}\ch{F}   &0\cr
  }\,,\text{ }\text{ }\text{ }\text{ } 
  \conslaw^{\ch{S}}=
 \kbordermatrix{
     & \cr
    \color{g}\ch{E}    &0\cr
    \color{g}\ch{EF}   &0\cr
    \color{g}\ch{EW}   &0\cr
    \color{g}\ch{E^{*}}   &1\cr
    \color{g}\ch{P} &1\cr
    \color{g}\ch{W}   &0\cr
    \color{g}\ch{S} &1\cr
    \color{g}\ch{F}   &0\cr
  }\,,\text{ }\text{ }\text{ }\text{ } 
 \conslaw^{\ch{F}}=
 \kbordermatrix{
     & \cr
    \color{g}\ch{E}    &0\cr
    \color{g}\ch{EF}   &1\cr
    \color{g}\ch{EW}   &1\cr
    \color{g}\ch{E^{*}}   &1\cr
    \color{g}\ch{P} &0\cr
    \color{g}\ch{W}   &1\cr
    \color{g}\ch{S} &0\cr
    \color{g}\ch{F}   &1\cr
  }\,,
 \label{eq:example_cl}
\end{equation}
and the concentrations of the corresponding moieties, i.e., the enzyme, the substrate and the fuel moiety:
$\consquantity^{\ch{E}} = [\ch{E}] + [\ch{EF}] + [\ch{EW}] + [\ch{E^{*}}] $, 
$\consquantity^{\ch{S}} = [\ch{E^{*}}] + [\ch{P}] + [\ch{S}] $,
$\consquantity^{\ch{F}} = [\ch{EF}] + [\ch{EW}] + [\ch{E^{*}}] + [\ch{W}] + [\ch{F}]$.

In the following, we examine how the different thermodynamic quantities introduced in the decomposition of the entropy production rate~\eqref{eq:epr_decomposition} change according to the control mechanism in some prototypical cases. 
We use $k_{+3}$, $k_{+3}/k_{+1}$, and $RT(k_{+3})^2/k_{+1}$ as units of measure for time, concentration, and thermodynamic quantities, respectively.
We assume that $k_{\pm1}=k_{-2}=k_{\pm3}=k_{\pm4}=k_{\pm5}=2$, $k_{+2}=2e$, 
$\stchempotential_{\ch{E}}= \stchempotential_{\ch{W}} = \stchempotential_{\ch{S}} = 1$, $\stchempotential_{\ch{EW}} = \stchempotential_{\ch{P}} = \stchempotential_{\ch{F}} = 2$, $\stchempotential_{\ch{EF}} = 3$, $\stchempotential_{\ch{E^{*}}} = 4$, 
and the initial condition is given by $[\ch{E}]=0.15$, $[\ch{EF}]=0.1$, $[\ch{EW}]=0.1$, $[\ch{E^{*}}]=0.1$, $[\ch{P}]=0.5$, $[\ch{W}]=0.5$, $[\ch{S}]=1$, $[\ch{F}]=1$.
The solution of the rate equation~\eqref{eq:rate}, where the currents satisfy mass-action kinetics~\eqref{eq:massaction}, has been numerically computed using the standard SciPy ODE solver and then employed to determine the thermodynamic quantities using their definitions. 

\subsection{Detailed Balanced Case}
We start by considering the case where only the species $\setexchanged = \{\ch{S},\ch{F}\}$ are exchanged with the surroundings.
The conservation laws $\conslaw^{\ch{S}}$ and $\conslaw^{\ch{F}}$ are broken and, consequently, 
i) both the species $\ch{S}$ and $\ch{F}$ are potential species $\setpotential$
and ii) the CRN is (unconditionally) detailed balanced.
According to Eq.~\eqref{eq:conc_moiety}, the concentrations of the exchanged moieties are given by the concentrations of the substrate and the fuel moiety,
\begin{equation}
\moieties=\begin{pmatrix}
\consquantity^{\ch{S}} \\ 
\consquantity^{\ch{F}}\\
\end{pmatrix}
\end{equation}
since $\matLbYp$ is the identity matrix in this particular case where $\matLb$, whose rows are the broken conservation laws, is speficied by
\begin{equation}
\matLb = 
 \kbordermatrix{
    & \color{g}\ch{E} &\color{g}\ch{EF}&\color{g}\ch{EW}&\color{g}\ch{E^{*}}&& \color{g}\ch{P}&\color{g}\ch{W}&&\color{g}\ch{S}&\color{g}\ch{F}\cr
    \color{g}\boldsymbol\conslaw^{\ch{S}} &0 &0& 0 & 1&\color{g}\vrule&1 &0& \color{g}\vrule&1 & 0\cr
    \color{g}\boldsymbol\conslaw^{\ch{F}} &0 &1& 1 & 1&\color{g}\vrule&0 &1& \color{g}\vrule&0 & 1\cr
  }\,.
\label{eq:example_matL}
\end{equation}

Thus, the semigrand free energy~\eqref{eq:semigrand} is given by
\begin{equation}
\semigrand = G -  \chempotential_{\ch{S}}\consquantity^{\ch{S}} - \chempotential_{\ch{F}}\consquantity^{\ch{F}}\,,
\label{eq:ex_semigrand}
\end{equation}
the driving work~\eqref{eq:cdwr} is specified as
\begin{equation}
\cdwr = -\Big(\ddt \chempotential_{\ch{S}} \Big) \consquantity^{\ch{S}} -\Big(\ddt \chempotential_{\ch{F}} \Big) \consquantity^{\ch{F}} \,.
\label{eq:ex_cdwr}
\end{equation}
while the nonconservative work~\eqref{eq:ncwr} vanishes because there are no $\setforce$ species.

\begin{figure*}[t]
  \centering
  \includegraphics[width=2.\columnwidth]{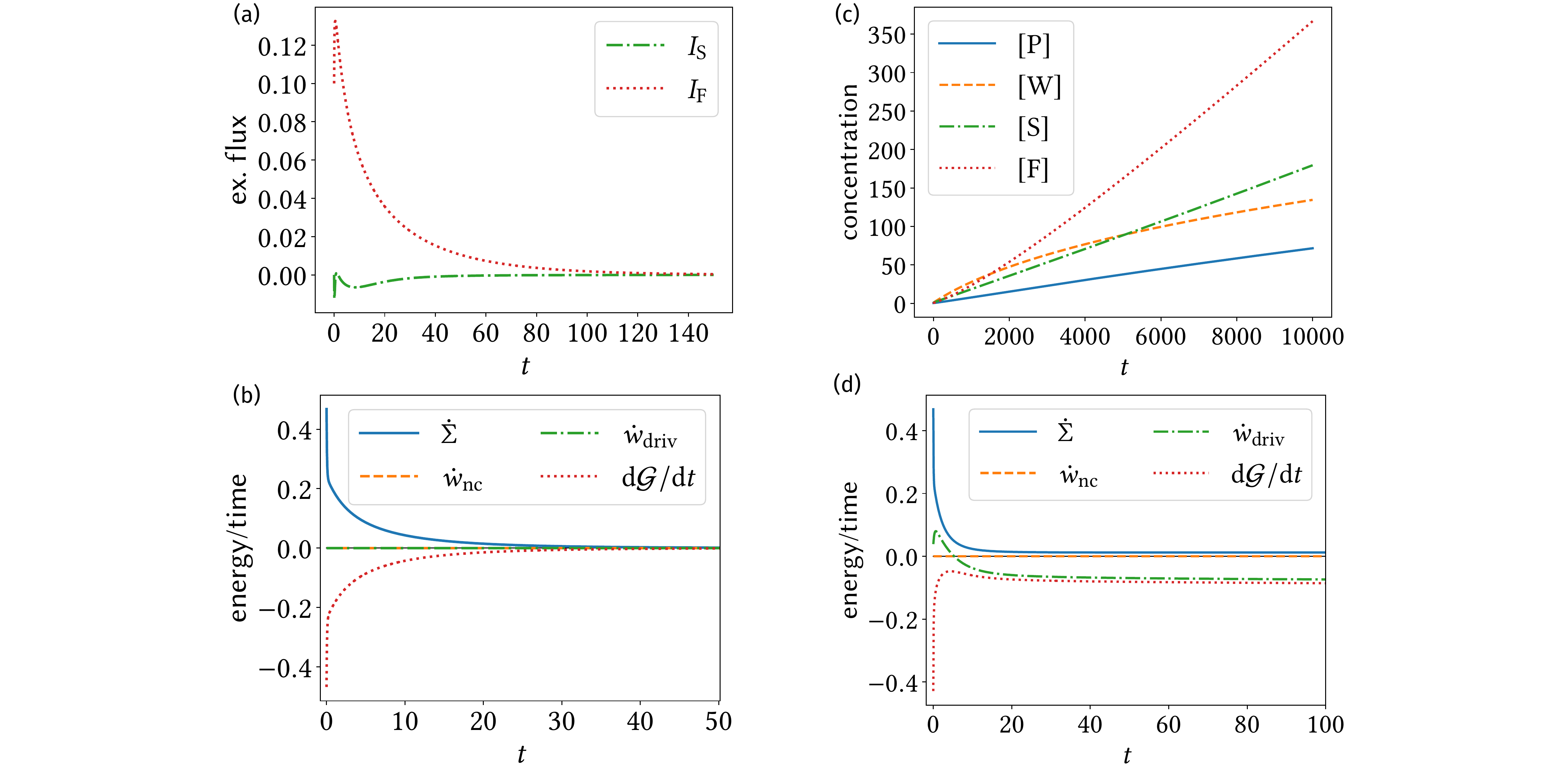}
\caption{
Exchange \exfluxes\text{ }(a) and thermodynamic quantities (b) of the CRN~\eqref{eq:example} when the species \ch{S} and \ch{F} are subjected to an autonomous concentration control.
Concentrations of \ch{P}, \ch{W}, \ch{S}, and \ch{F} (c),  and thermodynamic quantities (d) of the CRN~\eqref{eq:example} when the species \ch{S} and \ch{F} are subjected to an autonomous flux control with $\excurrel_{\ch{S}} = 0.025$ and $\excurrel_{\ch{P}} = 0.05$.}
\label{fig:db_case}
\end{figure*}
By applying an autonomous concentration control on the $\setexchanged$ species, the CRN will eventually reach an equilibrium steady state.
In the transient, the exchange \exfluxes\text{ }$\excurrel_{\ch{S}}$ and $\excurrel_{\ch{F}}$ balance the variation of the concentration $[\ch{S}]$ and $[\ch{F}]$ due to the chemical reactions (see Fig.~\ref{fig:db_case}a).
When equilibrium is reached, all the reaction currents vanish and hence also the exchange \exfluxes\text{ }(see Eq.~\eqref{eq:excurr_constconccontrol} and Fig.~\ref{fig:db_case}a).
From a thermodynamic stand point, the dynamics corresponds to a pure dissipation of the semigrand free energy as shown in Fig.~\ref{fig:db_case}b.
Indeed, no driving work is performed and Eq.~\eqref{eq:epr_decomposition} simplifies to $\dddt\semigrand = -T\epr\leq 0$.
 
On the other hand, by imposing an autonomous flux control that continuously injects $\ch{S}$ and $\ch{F}$ in the CRN at constant rate, namely $\excurrel_{\ch{S}}$ and $\excurrel_{\ch{F}}$ have constant positive values, the concentrations $[\ch{P}]$, $[\ch{W}]$, $[\ch{S}]$, and $[\ch{F}]$ continuously grow as shown in Fig.~\ref{fig:db_case}c.
The concentrations $[\ch{E}]$, $[\ch{EF}]$, $[\ch{EW}]$, and $[\ch{E^{*}}]$ are instead bounded since the concentration of the enzyme moiety $\consquantity^{\ch{E}}$ is still conserved (in other words, $\conslaw^{\ch{E}}$ is an unbroken conservation law). 
In this case, the dynamics never reaches a steady state.
From a thermodynamic standpoint, the control mechanism performs a driving work which maintains the CRN out of equilibrium (see. Fig.~\ref{fig:db_case}d).
In the long time limit, the driving work continuously extracts free energy at (almost) constant rate ($\cdwr\simeq-0.08$) which is provided by the constant decrease of the semigrand free energy ($\dddt\semigrand<\cdwr<0$). 
In parallel, the dissipation is very small ($0<\epr\ll|\dddt\semigrand|$) which implies that the CRN is constantly close to an equilibrium state that is changing in time because of the driving work.
Note that this is independent of the magnitude of $\excurrel_{\ch{S}}$ and $\excurrel_{\ch{F}}$.
In the long time limit, the concentrations of $\ch{S}$ and $\ch{F}$ are large enough that the exchange processes act as a slow perturbation
thus allowing the CRN to stay close to equilibrium.

\subsection{Nondetailed Balanced Case}
We consider now the case where also the species \ch{P} and \ch{W} are exchanged, i.e., $\setexchanged = \{\ch{P}, \ch{W}, \ch{S}, \ch{F}\}$.
Since no further conservation laws are broken, namely $\conslaw^{\ch{E}}$ is still an unbroken conservation law, the species \ch{P} and \ch{W} are force species $\setforce$.
Thus, the expressions of $\semigrand$ and $\cdwr$ are still the same as in Eq.~\eqref{eq:ex_semigrand} and~\eqref{eq:ex_cdwr},  respectively, while the nonconservative work is now given by
\begin{equation}
\ncwr = (\chempotential_{\ch{P}}-\chempotential_{\ch{S}})\excurrel_{\ch{P}} + (\chempotential_{\ch{W}}-\chempotential_{\ch{F}})\excurrel_{\ch{W}} \,,
\end{equation}
since also the matrix $\matLbYf$ is the identity matrix in this particular case given $\matLb$ in Eq.~\eqref{eq:example_matL}.

\begin{figure*}[t]
  \centering
  \includegraphics[width=2.\columnwidth]{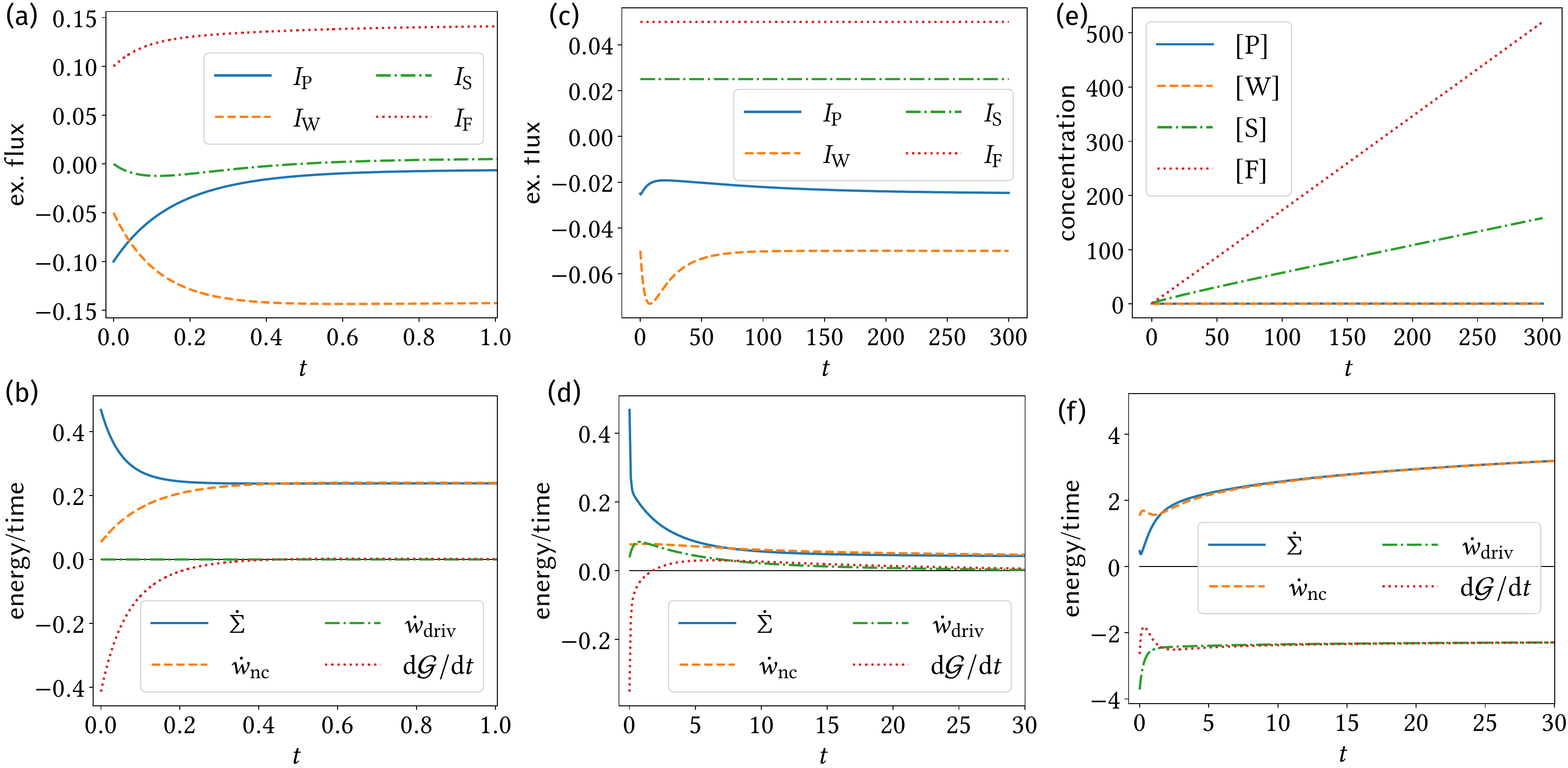}
\caption{Exchange \exfluxes\text{ }(a) and thermodynamic quantities (b) of the CRN~\eqref{eq:example} when the species \ch{P}, \ch{W}, \ch{S} and \ch{F} are subjected to an autonomous concentration control.
Exchange \exfluxes\text{ }(c) and thermodynamic quantities (d) of the CRN~\eqref{eq:example} when the species \ch{P}, \ch{W}, \ch{S} and \ch{F} are subjected to an autonomous flux control with $\excurrel_{\ch{P}}=-0.05[\ch{P}]$, $\excurrel_{\ch{W}}=-0.1[\ch{W}]$, $\excurrel_{\ch{S}}=0.025$, and $\excurrel_{\ch{F}}=0.05$.
Concentrations of \ch{P}, \ch{W}, \ch{S}, and \ch{F} (e), and thermodynamic quantities (f) of the CRN~\eqref{eq:example} when the species \ch{P}, \ch{W}, \ch{S} and \ch{F} are subjected to an autonomous flux control with $\excurrel_{\ch{P}}=-1[\ch{P}]$, $\excurrel_{\ch{W}}=-2[\ch{W}]$, $\excurrel_{\ch{S}}=1$, and $\excurrel_{\ch{F}}=2$.}
\label{fig:ndb_case}
\end{figure*}
When an autonomous concentration control maintains the concentrations of the species \ch{F}, \ch{S}, \ch{W}, and \ch{P} constant (like in Fig.~\ref{fig:chemostats}), the CRN relaxes towards a nonequilibrium steady state.
Indeed, the CRN continuously exchanges the $\setexchanged$ species with the surroundings (see the \exfluxes\text{ }$\excurrY$ in Fig.~\ref{fig:ndb_case}a).
At steady state, the species \ch{F} (resp. \ch{S}) is constantly injected into the CRN, i.e., $\steady{\excurrel}_{\ch{F}}>0$ (resp. $\steady{\excurrel}_{\ch{S}}> 0$), while the species \ch{W} (resp. \ch{P}) is constantly extracted at the same rate, i.e., $\steady{\excurrel}_{\ch{W}} = - \steady{\excurrel}_{\ch{F}} <0$ (resp. $\steady{\excurrel}_{\ch{P}} = - \steady{\excurrel}_{\ch{S}}<0$).
This physically means that the CRN operates as a steady state engine interconverting $\ch{F}$ into $\ch{W}$ and $\ch{S}$ into $\ch{P}$. 
From a thermodynamic standpoint, the dynamics continuously dissipates (free) energy (see Fig.~\ref{fig:ndb_case}b).
In the transient, the dissipated energy $\epr$ is provided by the decrease of the semigrand free energy $\dddt\semigrand<0$, and by the nonconservative work  $\ncwr>0$.
At steady state, the dissipation due to the continuous interconversion of $\ch{F}$ into $\ch{W}$ and $\ch{S}$ into $\ch{P}$ is solely balanced by the nonconservative work, i.e., $\steady{\epr} = \ssncwr$, since the semigrand free energy is constant, i.e., $\dddt\semigrand =0$.

We now turn to the situation where an autonomous flux control is applied to the CRN.
As in Fig.~\ref{fig:processes}, we consider that the species \ch{S} and \ch{F} are continuously injected in the CRN at constant rates $\excurrel_{\ch{S}}$ and $\excurrel_{\ch{F}}$, while the species \ch{P} and \ch{W} are extracted at rates that are proportional to their concentrations, i.e., $\excurrel_{\ch{P}} = -k_{\ch{P}}[\ch{P}]$ and $\excurrel_{\ch{W}} = -k_{\ch{W}} [\ch{W}]$.
Depending on the injection fluxes $\excurrel_{\ch{S}}$ and $\excurrel_{\ch{F}}$ and the extraction rate constants $k_{\ch{P}}$ and $k_{\ch{W}}$, the dynamics can be significantly different as shown in Fig.~\ref{fig:ndb_case}.

When for example $\excurrel_{\ch{P}}=-0.05[\ch{P}]$, $\excurrel_{\ch{W}}=-0.1[\ch{W}]$, $\excurrel_{\ch{S}}=0.025$, and $\excurrel_{\ch{F}}=0.05$, the CRN relaxes towards a steady state like when the concentration control is applied.
At steady state the extraction fluxes of the species ${\ch{P}}$ and ${\ch{W}}$ balance exactly the injection fluxes of the species ${\ch{S}}$ and ${\ch{F}}$ (Fig.~\ref{fig:ndb_case}c): $\steady{\excurrel}_{\ch{W}} = - \steady{\excurrel}_{\ch{F}}$ and $\steady{\excurrel}_{\ch{P}} = - \steady{\excurrel}_{\ch{S}}$.
 The CRN operates again as a steady state engine interconverting $\ch{F}$ into $\ch{W}$ and $\ch{S}$ into $\ch{P}$. 
From a thermodynamic standpoint (see Fig.~\ref{fig:ndb_case}d), we identify three different regimes.
For $t<1.7$, the energy dissipated by the dynamics $\epr$ is provided by the decrease of the semigrand free energy $\dddt\semigrand<0$, by the nonconservative work $\ncwr>0$ and also by the driving work $\cdwr>0$.
For $1.7 < t < 26$, the energy provided by the nonconservative work and the driving work is greater than the dissipation, $\ncwr+\cdwr-\epr>0$, leading to an increase of the semigrand free energy $\dddt\semigrand>0$.
For $t > 26$, steady state is reached: the driving work as well as the time derivative of the semigrand free energy vanish and only the nonconservative work balances the dissipation, i.e., $\steady{\epr} = \ssncwr$, like in the case of concentration control.
Thus, the steady state dynamics and thermodynamics under flux control (unlike the transient ones) are qualitatively the same as under concentration control.
However, the specific steady state to which the CRN relaxes depends control mechanism.
This can be easily verified by comparing the values of the exchange fluxes in Fig.~\ref{fig:ndb_case}a and~\ref{fig:ndb_case}c as well as of the thermodynamic quantities in Fig.~\ref{fig:ndb_case}b and~\ref{fig:ndb_case}d.

When for example $\excurrel_{\ch{P}}=-1[\ch{P}]$, $\excurrel_{\ch{W}}=-2[\ch{W}]$, $\excurrel_{\ch{S}}=1$, and $\excurrel_{\ch{F}}=2$,
the CRN never reaches steady state.
The concentrations of \ch{S} and \ch{F} continuously grow (Fig.~\ref{fig:ndb_case}e), while the concentrations of \ch{P} and \ch{W} become constant and very small.
This happens because the species \ch{S} and \ch{F} are injected too quickly into the CRN to be effectively interconverted into \ch{P} and \ch{W} and extracted: $\excurrel_{\ch{F}} > -\excurrel_{\ch{W}} $ and $\excurrel_{\ch{S}} > -\excurrel_{\ch{P}} $. 
Thus, they accumulate. 
From a thermodynamic standpoint (see Fig.~\ref{fig:ndb_case}f), the dissipation continuously increases and it is mostly balanced by the nonconservative work, i.e., $\epr \simeq \ncwr$.
On the other hand, the driving work continuously extracts energy from the CRN which is provided by the decrease of the semigrand free energy, i.e., $\cdwr \simeq -\dddt{\semigrand}$.
This thermodynamic behavior is not specific to the model: it is a direct consequence of the CRN being close to a nonequilibrium steady state that changes in time.
Indeed, in the long time limit the concentrations of $\ch{S}$, $\ch{F}$ are large enough that the exchange processes act as a perturbation, while the concentrations \ch{P} and \ch{W} are almost constant. 
This implies that the entropy production rate is (mostly) balanced by the non conservative work (as if the CRN was at steady state), while the energetic cost of changing the steady state accounted by $\dddt{\semigrand}$ is provided by the driving work.
In this regime, the dynamics and thermodynamics under flux control are significantly different from those under concentration control.

\section{Discussion\label{sec:discussion}}
We discuss now the general differences between concentration and flux control focusing in particular on the various contributions to the entropy production rate~\eqref{eq:epr_decomposition}.
In the initial transient of the dynamics, the flux control always leads to a time dependent variation of the concentrations of the exchanged species which can be obtained only with an nonautonomous concentration control.
This happens because the exchange fluxes do not balance the variation to the concentrations due to the chemical reactions. 
Thus, part of the free energy exchange between CRNs and the surroundings is always given by a driving work contribution if the flux control is applied.
If CRNs are nondetailed balanced, another contribution to the free energy exchange comes from the nonconservative work independently of whether concentration control or flux control is imposed.

In the long time limit, CRNs show (qualitatively) the same phenomenology with both control mechanisms if a steady state is reached: the dissipation is balanced by the nonconservative work, i.e., $\steady{\epr} = {\ssncwr}$.
On the other hand, the relaxation towards a steady state is not always granted under flux control. 
When this is the case, we found 
that the variation to the semigrand free energy of CRNs is (almost) balanced by the driving work, i.e., $\dddt\semigrand \simeq \cdwr$, while the dissipated free energy (in case of nondetailed balanced CRNs) is (mostly) provided by the nonconservative work, i.e., $\epr \simeq \ncwr$.
This is a consequence of the fact that CRNs evolve close to a steady state which changes in time.
If CRNs are detailed balanced, the entropy production rate becomes very small compared to the absolute value of the variation of the semigrand free energy, i.e., $\epr\ll |\dddt\semigrand|$, because CRNs evolve close to an equilibrium steady state.

Our results have been obtained for ideal dilute solutions, but they can be easily generalized to non-ideal solutions by recognizing that the chemostats directly control the chemical potentials of the exchanged species via the chemical reaction~\eqref{eq:ex_process_chemostat} instead of their concentrations~\cite{Avanzini2021}.

Our work opens the way to the investigation of the energetic cost of growth in autocatalytic processes.
Using kinetic arguments, mathematicians have proven that for a specific type of CRNs (called single linkage class CRNs) continuous growth is prohibited under concentration control~\cite{Anderson2011, Anderson2011b}. 
According to preliminary numerical observations, our thermodynamic approach indicates that the concentration control cannot provide enough energy to balance both the dissipation and the increase of the semigrand free energy due to continuous autocatalytic growth.
However, we find that flux control might maintain the CRN close to a nonequilibrium steady state where the concentrations of the autocatalytic species continuously increase so powering continuous growth.
We leave the rigorous characterization of these numerical observations to future studies.

\section{Acknowledgments}
This research was funded by the European Research Council project NanoThermo (ERC-2015-CoG Agreement No.~681456). 
We thank the reviewers for their valuable comments.

\bibliography{biblio}
\end{document}